\documentclass[a4paper,11pt]{article}
\usepackage{pos}

\title{Staggered bosons and Kahler-Dirac bosons}

\author*[a,b,c]{David Berenstein}
\author[d]{Simon Catterall}
\author[a]{P.N. Thomas Lloyd}

\affiliation[a]{Department of Physics, University of California\\
  Santa Barbara, CA 93106, U.S.A.}

\affiliation[b]{Institute of Physics, University of Amsterdam, Science Park 904,
PO Box 94485, 1090 GL Amsterdam, The Netherlands}

\affiliation[c]{Delta Institute for Theoretical Physics, Science Park 904,
PO Box 94485, 1090 GL Amsterdam, The Netherlands}

\affiliation[d]{Syracuse University, Department of Physics, Syracuse, 13244, New York, USA}

\emailAdd{dberens@physics.ucsb.edu}
\emailAdd{smcatterall@gmail.com}
\emailAdd{plloyd@ucsb.edu}

\abstract{We describe a novel way to think about bosonic lattice theories in Hamiltonian form where each lattice site has only a half boson degree of freedom. The construction requires a non-trivial Poisson bracket between neighboring sites and leads to gapless theories with non-invertible symmetries. We also describe a bosonic version of Kahler-Dirac fermions, dubbed Kahler-Dirac bosons that can be performed on any triangulation of a manifold. This also leads to a straightforward implementation of supersymmetry on the lattice and one immediately deduces the Dirac equation of the corresponding Kahler-Dirac fermions.}

\FullConference{Corfu Summer Institute 2023 "School and Workshops on Elementary Particle Physics and Gravity" (CORFU2023)\\
 23 April - 6 May , and 27 August - 1 October, 2023\\
Corfu, Greece\\}

\begin{document}
\maketitle

\section{Introduction}

From the point of view of Hamiltonian physics, position and momentum variables appear on the same footing. That is, there is a canonical transformation exchanging position and momenta. Such possible symmetries are not apparent in a Lagrangian formulation of a quantum field theory. One can also notice that the Hamiltonian equations  of motion are first order in time and in that sense they can be similar to a Dirac equation. 

When one considers staggered fermions, they mitigate the doubling problem on the lattice \cite{Kogut:1974ag}. Roughly speaking, in a staggered fermion formulation the degrees of freedom of a fermion are distributed on different lattice sites: they eventually combine between various neighbors to produce the continuum degrees of freedom. In $1+1$ dimensions, a Majorana fermion chain in the continuum limit produces a Dirac fermion: a left mover and a right mover. A Majorana fermion has only one fermionic variable degree of freedom per lattice site 
$\theta_j$ satisfying $\theta_j^2=1$. In that sense, it is a half fermion. 
 
It is interesting to ask if  one can find bosonic theories on the lattice with similar characteristics.  Such a theory will naturally mimic properties of  staggered fermions and the name staggered boson is an apt description of such systems. One can also call them a theory of a half boson degree of freedom.

In this paper, we will show such a construction in $1+1$ dimensions. Once one understands that it is possible to mimic the physics of a staggered 1 dimensional fermion, one can ask if more general classes of discretized fermions are possible. In particular, if one wants to work on curved manifolds, the natural lattice fermion system that arises is the so-called Kahler-Dirac fermions. We will show that there is a similar construction of a Kahler-Dirac boson which makes the combination of a Kahler-Dirac boson plus a Kahler-Dirac fermion
natural for supersymmetry: there is a simple supersymmetry that squares to the Kahler-Dirac form for the fermion hamiltonian.  The results concerning the Kahler-Dirac bosons and their relation to supersymmetry are preliminary. They are currently being investigated in joint work by  by D. Berenstein and S. Catterall.

\section{The chiral boson}
Before giving examples in the lattice, one can motivate the idea of a half bosons degree of freedom  by starting with a chiral boson. The chiral boson is a theory where a bosonic field satisfies a first order differential wave equation.
For example, consider a field $\phi(x,t)$ that satisfies
\begin{equation}
\partial_t \phi = c \partial_x \phi ,\label{eq:chiral}
\end{equation}
where $c$ is the speed of light, 
will also satisfy the following equation
\begin{equation}
\partial_t^2 \phi= c^2 \partial_x^2\phi. \label{eq:full}
\end{equation}
In that sense, the chiral boson has a first order differential equation rather than second order. It is a square root of the wave equation. Also, if one quantizes the field $\phi$
 into modes,  only half of the oscillators of a full bosonic degree of freedom are present. 

We can now ask if there is a Hamiltonian description of the chiral boson degrees of freedom \eqref{eq:chiral}, which involve a non-trivial Poisson bracket structure for the $\phi$ degrees of freedom and which reproduce the first order equation without additional projections  of the general solution of \eqref{eq:full}.
The answer is yes. One introduces a non-trivial Poisson bracket
\begin{equation}
\{ \phi(x), \phi(x')\}_{P.B.}= \partial_x \delta(x-x')\label{eq:KM}
\end{equation}
and the Hamiltonian function
\begin{equation}
H= c \int dx  \frac{\phi^2}2\label{eq:sugawara}
\end{equation}
It is easy to check that these two together reproduce \eqref{eq:chiral}. Notice that in the Poisson bracket, there is not an obvious canonical conjugate momentum variable to $\phi$, which one would usually call $\pi$, which in many setups one wold also identify with $\dot \phi$ in the Lagrangian formulation.
The Poisson bracket is instead an antisymmetric constant distribution and on its own defines a notion of phase space.
If one changes labels and calls $\phi\equiv J(x,t)$, the commutator equations can be more readily recognized as the Lorentzian version of the left-moving $U(1)$ Kac-Moody current algebra. The derivative of the delta function is the Schwinger anomaly, and the Hamiltonian \eqref{eq:sugawara} is in the Sugawara form .

Our goal in defining staggered bosons is to have a Poisson bracket that produces a discretized form of \eqref{eq:KM}, and to copy the Hamiltonian construction of \eqref{eq:sugawara}.
What will be interesting is that these two ingredients together produce a gapless theory in the continuum theory and that the theory built this way has an enhanced symmetry. This symmetry will be a non-invertible symmetry that will play an important role in the dynamics.

\section{Staggered bosons}

The idea of staggered bosons is to assign a single bosonic variable to each lattice site $q_j$.  For simplicity we will work in a one dimensional lattice system.
Unlike regular bosons where one would assign  a canonical conjugate pair $x_j, p_j$ to a lattice site, we have instead only one bosonic variable. For any such variable, a fudamental property of Poisson brackets is that all variables commute with themselves  $\{q,q\}_{P.B.}=0$.
If we want a non-trivial equation of motion, we need a non-vanishing Poisson bracket. This means that it must be the case that $\{q_j, q_k\}_{P.B.}\neq 0$ for some $j,k$.
We also want it to be translation invariant and similar in form to \eqref{eq:KM}. The simplest such bracket that is approximately local, translation invariant and a constant is given by
\begin{equation}
\{q_j,q_k\}_{P.B.}= \delta_{j,k-1}-\delta_{j,k+1} .\label{eq:PB}
\end{equation}
This is a discretized version of the derivative of the delta function. It cannot be exactly local because it needs to be antisymmetric.
The Poisson bracket is non-trivial just between nearest neighbors. 
Abstractly, the phase space and quantum algebra defined by \eqref{eq:PB} with the additional requirement of having locality of the Hamiltonian is the staggered boson.
This construction was done in \cite{Berenstein:2023tru}.

If we choose  the Hamiltonian given by
\begin{equation}
H= c \sum_j \frac{q_j^2}{2}
\end{equation}
one can easily see that
\begin{equation}
\dot q_j= q_{j+1}-q_{j-1}
\end{equation}
gives a discretized version of the one dimensional wave equation. The Poisson bracket \eqref{eq:PB} is providing the coupling to the neighboring sites.
In the continuum limit, the equation of motion becomes relativistic and corresponds to a massless particle. 

Notice also that the hamiltonian was the simplest one we could consider, so the gaplessness of the continuum limit arose without fine tuning the Hamiltonian. 
 The simplest way to understand the physics is to make use of translation invariance and the linearity of the equations of motion. Using a plane wave form of $q_j\simeq \exp(i k j-i\omega(k) t)$, one readily finds a dispersion relation 
 \begin{equation}
 -i \omega(k) = \exp(i k)- \exp(-i k)= 2 i \sin(k)
 \end{equation}
This is a periodic function of the quasimomentum $k$. The sign that we have a first order differential equation is that $\omega(k)$ is a single valued function of $k$, rather than a square root. There are two types of low frequency modes. Those at quasi-momentum close to zero $k\simeq 0$, and those at quasimomomentum $k\simeq \pi$. The first set are left movers, and the other set are right moving degrees of freedom. 

The appearance of right movers should not be surprising. A pure left mover has a gravitational anomaly $c_L=1$, but the lattice theory has not gravitational anomaly in the UV. If there is a contribution from left movers, there must be a similar contribution to the right movers to cancel the anomaly $a= c_L-c_R=0$. Thus, even though it seemed as if we were getting only a left mover, the theory also produces a right mover.  Basically, these additional massless modes are doublers. In this sense, this is a different incarnation  of the Nielsen-Ninomiya theorem \cite{Nielsen:1981hk}.

The quantization condition requires that the mode at quasimomentum $k$ is conjugate to the mode at quasimomentum $-k$. Because of this property $\omega(-k)= -\omega(k)$.
This always produces a crossing of zero at $k=0,\pi$, so the gapless modes are robust. This is true even in the presence of translation invariant small perturbative corrections, which would preserve this property for quasi-particles.

\subsection{Non-invertible symmetries}

The algebra of the $q_j$ has a natural automorphism $q_j\to q_{j+1}$ of translations. It turns out that this is not necessarily a symmetry of the Hilbert space representations.
There are two natural central elements of the algebra $C_{even}= \sum_j q_{2j}$ and $C_{odd}= \sum_j q_{2j+1}$. It is easy to see that these commute with all the $q_j$.
Both of these survive if we have an even number of sites on the lattice with periodic boundary conditions. 
The other degrees of freedom are given by copies of Harmonic oscillators (the  Fourier modes at $k\neq 0,\pi$). Each of these is a copy of the Weyl algebra.

Because $C_{even}, C_{odd}$ are central,  they act as $c$-numbers on irreducible representations of the algebra of the $q$.
 The Hilbert space states are classified by
these numbers. The translation symmetry of the algebra $q_j\to q_{j+1}$ acts by exchanging these two with each other. 
This is an outer-automorphism of the algebra, which we will call ${\cal D}$. 

For the oscillator modes, the Stone-von Neumann theorem guarantees that the automorphism of the Weyl-algebra can be lifted to a unitary action on the Hilbert space. 
On the other hand, for the central elements there is no such guarantee. Translation by two restores the central elements and can be lifted to a unitary of the full Hilbert space.
We call this translation $T$. 

There should be a sense in which ${\cal D}$ is a symmetry on the Hilbert space of the theory. Naively, we would want to state that the translation by $2$ which is a godd symmetry, is roughly ${\cal D}$ acting twice. This is true at the level of the automorphism of the algebra of the $q$, so we want a sense in which  ${\cal D}^2 \simeq T$. The point is that the notion of ${\cal D}$ on the Hilbert space is slightly inconsistent. What we can do instead is to consider a projection $\Pi$ onto pairs of labels of the states $(C_{odd},{C_{even}})$ that are paired (can be swapped), that is labels for which the mirror pairs  $(C_1,C_2)$ and $(C_2,C_1)$ are both allowed. On any other state, we define $\Pi$ as acting by zero.
Given the full Hilbert space ${\cal H}$, we can build a small Hilbert space ${\cal H}_s= \Pi{\cal H}$. The action of ${\cal D}$ (we can think of it as ${\cal D}_s$)  is well defined on ${\cal H}_s \subset {\cal H} $. 
We build an operator  ${\cal D}= {\cal D}_s \circ \Pi$ and one can readily check that 
\begin{equation}
{\cal D}^2= T\circ \Pi=\Pi \circ T
\end{equation}
This shows that the staggered boson has a non-invertible symmetry, characterized by ${\cal D}$ \cite{Berenstein:2023ric}.
It is this symmetry that is secretly protecting the gaplessness of the system. 

On configurations where it acts non-trivially, the non-invertible symmery ${\cal D}$ is similar to T-duality. This requires a precise understanding of how the staggered boson degrees of freedom can be described  starting from a  regular boson on the lattice \cite{Berenstein:2023tru}. A more general account of how non-invertible symmetries 
control anomalies in lattice systems can be found in \cite{Cheng:2022sgb}.

One can also consider more general algebras of operators on lattices. For example, we can ask that at each lattice site there is a  discrete (unitary) variable $K_j$ such that $K_j^N=1$ for some $N$ and such that 
\begin{equation}
 K_j K_k = \eta^{ \delta_{j,k-1}-\delta_{j,k+1}} K_k K_j, \label{eq:ncphase}
 \end{equation}
  where $\eta$ is a (primitive) N-th  root of unity. This is a multiplicative version
of a non-trivial commutation relation between nearest neighbors.

With the Hamiltonian given by
\begin{equation}
H= -\sum_j K_j+k_j^{-1}
\end{equation}
one can check that the system has a similar non-invertible symmetry with ${\cal D}^2 = T\circ \Pi$ where $\pi$ is a projector onto a ${\mathbb Z}_N$ singlet.
These Hamiltonian lead to critical spin chains. When $N=2$, this is a rewriting of the Ising model in a transverse magnetic field at criticality. 
When $N=3$, this is a conformal field theory with central charge $c=4/5$. At $N=4$ one gets a CFT with central charge $c=1$, which is two copies of the Ising model at criticality.  
For $N>4$ the theory in the continuum limit is a $c=1$ boson with a $U(1)$ current algebra at radius $R=\sqrt{2N}$, and the non-invertible symmetry is the Kramers-Wannier duality  \cite{Berenstein:2023ric}. These non-invertible symmetries are qualitatively very similar to those found in Majorana chains \cite{Seiberg:2023cdc}, which are equivalent to the special case $N=2$ under the right identifications. The non-invertible translation of the Majorana chain, is equivalent to the non invertible ${\cal D}.$

\section{Kahler-Dirac bosons}

Since  bosonic theories based on \eqref{eq:PB} or \eqref{eq:ncphase} are very similar to Majorana fermions, it is instructive to ask if we can take a system that utilizes both of them and marries them into a supersymmetric system in a natural way.
The answer is yes. The most natural setup for such an equivalence is the one associated to Kahler-Dirac fermions.

The basic idea is that one can consider the Kahler-Dirac operator 
\begin{equation}
K= d-\delta
\end{equation}
where $d^2=0$ is the differential operation, and $\delta$ is its adjoint. This is well defined on any smooth manifold where it acts on the set of differential forms (of arbitrary degree).
It is easy to show that $ K K^\dagger= - \square$ is a formal square root of the (Euclidean) Laplacian. 
A natural Dirac equation for an equation of motion for such a fermion system is given by
\begin{equation}
\dot \psi \propto K \psi
\end{equation}
The idea to turn $K$ into a lattice operator is to use a map $d\to \bar d$, $\delta \to \bar \delta$ that acts naturally on discrete spaces. In this setup $\bar d$ is the coboundary operator and $\bar \delta= \bar d ^T$ is the boundary operator. The natural obejcts to consider are triangulations, or more generally $CW$ complexes on which one can easily defined homology and cohomology (see \cite{Catterall:2018dns,Catterall:2023nww} and references therein). Armed with such discrete operations, one naturally has $\bar d^2=0$ etc.

To make this a bit more precise, a finite triangulation ( or more generally a finite cellulation) is characterized by cells of various dimension $n$, which we will collectively call $C^{(n)}$. The number of these cells is finite and the individual cells of this dimension are labeled by extra index labels $j$.
The maps $\delta$ take cells of dimension $n$ to their boundary (a collection of cells of dimension $n-1$) as follows
\begin{equation}
\delta C_j^{(n)} = \sum c_{jk } C^{(n-1)}_k
\end{equation}
where the $C_j^{(n)}$ are all the possible cells of dimension $n$. The boundary of $C_j^{(n)}$ is characterized by the $c_{jk}$ which are signed integers. They are given by the number of oriented copies of $C^{(n-1)}_k$ on the boundary of $C_j^{(n)}$. 
Triangulations, or more general CW complexes are tools that allow one to find discrete approximations to arbitrary manifolds. It is usually possible after refining a 
decomposition, to choose the $c_{jk }$ so that they are $\pm1,0$.

The idea to build a Kahler-Dirac boson is to use the $c_{jk}$ as a building block for the Poisson bracket. First, to each $C_j^{(n)} $ we assign a bosonic variable $q_j^{(n)}$.
 We build a Poisson bracket by declaring that 
 \begin{equation}
 \{q_j^{(n)},q_k^{(n-1)}\}_{P.B.}= c_{jk}\label{eq:KDB}
 \end{equation}
and that any other pair of variables whose Poisson bracket is not already defined by this equation vanishes exactly. in this way variables associated to even cells will always commute with each other, and variables associated to odd cells will always commute with each other.
The Kahler-Dirac fermions will similarly attach a Grassman Majorana variable $\theta^{(n)}_j$ to each such cell.
Equivalently, we rephrase the Poisson bracket as canonical commutators
 \begin{equation}
 \ [q_j^{(n)},q_k^{(n-1)} ] = i \hbar c_{jk}\label{eq:KDBc}
 \end{equation}

Consider now the supersymmetry operator
\begin{equation}
Q= \sum_{n,j}  \frac{q_j^{(n)} \theta^{(n)}_j}{\sqrt{2}}
\end{equation}
If we compute $Q^2$, we get that there are two types of terms. The ones where the $\theta$ are the same, and the one where they are different.
These first are easy. One finds a contribution to the Hamiltonian that is of the form
\begin{equation}
H_q= \sum_{j,n} \frac{(q_j^{(n)})^2}{2}
\end{equation}

Let us study the case where the two fermion variables  are different $ \theta^{(n)}_j \theta^{(\tilde n)}_k$. There are two terms that contribute:
\begin{equation}
\theta^{(n)}_j q_j ^{(n)} \theta^{(\tilde n)}_k q^{(\tilde n)}_k+\theta^{(\tilde n)}_k q^{(\tilde n)}_k \theta^{(n)}_j q_j ^{(n)} ,
\end{equation}
but because the $\theta $ anticommute, we find that the term with this combination of fermion variables is sensitive to the commutator of the $q$ variables.
After a bit of effort, one finds that the Fermionic Hamiltonian looks like
\begin{equation}
H_f\simeq \sum i \theta_j^{(n)} (\delta  \theta_j^{(n)} )
\end{equation}
where $(\delta  \theta_j^{(n)} )= \sum_k  c_{jk } \theta^{(n-1)}_k$

This has the exact form that one would expect the Kahler-Dirac operator to take.

In the special case of a one dimensional lattice, we would have zero cells (even sites) and one cells (odd sites). The boundary operator of the one cells
$\delta C_{2j+1} = C_{2j+2}-C_{2j}$ would give the signed sum of the neighboring even sites. It is easy to check that this gives the exact same answer as 
\eqref{eq:PB}. In that sense the staggered boson is the same as a one dimensional Kahler-Dirac boson. The same statement is true for the staggered fermion.
The combination of the Kahler-Dirac boson plus the Kahler-Dirac fermion becomes equivalent to a staggered boson plus a Majorana chain (also known as a staggered fermion).
They have a natural supersymmetry relating them. 

It is interesting to understand better these constructions. Such work is being carried out by D. Berenstein and S. Catterall. 

\acknowledgments

Research supported in part by the Department of Energy under Award No. DE- SC0019139. 
S.C. is also supported by the Department of Energy under award DE-SC0009998.
DB is also supported in part by the Delta ITP consortium, a program of the Netherlands Organisation for Scientific Research (NWO) funded by the Dutch Ministry of Education, Culture and Science (OCW)

\end{document}